\documentclass[aps,prl,floats,twocolumn,showpacs,superscriptaddress]{revtex4}
 \usepackage{graphics}
 \usepackage{epsfig}
\usepackage{amsmath}
\usepackage{amssymb}
\begin{document}
\title{Spin wave excitations: The main source of the temperature dependence
of interlayer exchange coupling in nanostructures}
\author{S. Schwieger}
\affiliation{Technische Universit\"at Ilmenau, Theoretische Physik I,
           Postfach 10 05 65, 98684 Ilmenau, Germany}
\affiliation{Lehrstuhl Festk{\"o}rpertheorie, Institut f{\"u}r Physik,
Humboldt-Universit{\"a}t zu Berlin,
  Newtonstr.~15, 12489 Berlin, Germany}

\author{J. Kienert}
\email[Corresponding author, ]{kienert@physik.hu-berlin.de}
\affiliation{Lehrstuhl Festk{\"o}rpertheorie, Institut f{\"u}r Physik,
Humboldt-Universit{\"a}t zu Berlin,
  Newtonstr.~15, 12489 Berlin, Germany}

\author{K. Lenz}
\affiliation{Institut f\"ur Experimentalphysik, Freie Universit\"at 
Berlin, Arnimallee 14, 14195 Berlin, Germany}

\author{J. Lindner}
\altaffiliation[new address: ]{Fachbereich Physik, Experimentalphysik, 
Universit\"at Duisburg-Essen, Lotharstr.~1, 47048 Duisburg, Germany}

\affiliation{Institut f\"ur Experimentalphysik, Freie Universit\"at 
Berlin, Arnimallee 14, 14195 Berlin, Germany}

\author{K. Baberschke}
\affiliation{Institut f\"ur Experimentalphysik, Freie Universit\"at 
Berlin, Arnimallee 14, 14195 Berlin, Germany}

\author{W. Nolting}
\affiliation{Lehrstuhl Festk{\"o}rpertheorie, Institut f{\"u}r Physik,
Humboldt-Universit{\"a}t zu Berlin,
  Newtonstr.~15, 12489 Berlin, Germany}

\begin{abstract}
Quantum mechanical calculations based on an extended Heisenberg model are
compared with ferromagnetic resonance (FMR) experiments on
prototype trilayer systems Ni$_7$/Cu$_n$/Co$_2$/Cu(001) in order to
determine and separate for the first time quantitatively the sources of
the temperature
dependence of interlayer exchange coupling. Magnon excitations are
responsible for about 75$\%$ of the reduction of the coupling strength
from zero to room temperature. The remaining 25$\%$ are due to temperature effects
in the effective quantum well and the spacer/magnet interfaces. 
\end{abstract}
\maketitle

The coupling of two magnetic layers through a nonmagnetic spacer layer
(interlayer exchange coupling, IEC) has received considerable
attention recently. The physics at $T$=0 can be described quite
generally in the framework of a quantum interference model
\cite{Bruno95,Ed91,Stil05},
e.g. the dependence on the spacer thickness.
Within this theory, the period, the amplitude, and the phase of the coupling are well
understood today. However, most of the experiments on magnetic multilayers and their IEC have been undertaken at ambient temperature, 
not only for easier feasibility but also in view of technological applications. In this Letter we evidently show in theory and experiment 
that the major effect is due to magnon excitations, whereas band structure effects are less important.

Much experimental and theoretical work focused on the oscillating
dependence of the IEC on the spacer thickness resulting in ferromagnetic
(FM) and
antiferromagnetic (AFM) coupling of the magnetic layers \cite{UCP96}. As
for the temperature dependence there is convincing experimental evidence
on several
IEC systems that the coupling strength decreases with $T$ from its
$T$=$0$-value to almost zero at $T=T_\text{C}$ in metallic systems \cite{LM02,Zhang94,Pedi97}
and that it shows an effective $T^{3/2}$-dependence \cite{LM02}.
However relatively little theoretical work has been carried out concerning
the temperature dependence of the IEC \emph{over the full temperature
range}. The
detailed \emph{ab-initio} investigations which focused on Fermi edge
softening effects are restricted to low temperatures by the assumption
that the spin-resolved band
structure does not change significantly \cite{Bruno99,Albu96}. Working at
all temperatures, Mills and coworkers investigated the influence of spin 
waves and predicted the experimentally found $T^{3/2}$-dependence
\cite{Mills94,LM02}.

In this Letter we present a novel approach which combines a quantum
mechanical treatment of magnetic multilayers at all temperatures with
experiments on
the prototype IEC systems Ni$_7$/Cu$_n$/Co$_2$/Cu(001) (subscript: number
of layers).
Up to now a quantitative separation of the possible sources of the
$T$-dependence (spacer effects like softening of the Fermi surface or
reduction of the spin asymmetry
of the reflection coefficients vs. magnetic layer effect, i.e.\ excitation
of magnons) has not been achieved \cite{SN04}.
The key benefit of the current approach is the possibility to ``switch on
and off'' magnon excitations in our model. Thus, by performing
one-parameter fits of the
microscopic interlayer exchange coupling with and without magnons, we can
separate the two main sources of the temperature
dependence of the IEC. We find that the magnetic contribution exceeds the
spacer part by a factor of 3 in the range from $\approx 0.15\cdot
T_\text{C}$ to $T_\text{C}$.

The theoretical description of $T$-dependent IEC experiments often relies
on the classical, macroscopic Landau-Lifshitz (LL) equations. This
continuum's approach
is usually used together with an expansion of the free energy to obtain
the IEC energy by defining
\cite{SN04,LB03}
\begin{equation}
2J_\text{ inter}(T) = F_\text{IEC}(T) = F_{\uparrow\uparrow}(T) -
F_{\uparrow\downarrow}(T)\;.
\label{Fiec}
\end{equation}
Although the LL approach allows for an examination of the overall
$T$-dependence of the IEC, it cannot distinguish between the different
contributions, i.e.,
it cannot yield (\ref{Fiec}) in the form $F_\text{IEC}(T) =
F_\text{spacer}(T) + F_\text{magn.}(T)$.

\begin{figure}[t]
\includegraphics[width=0.8\linewidth]{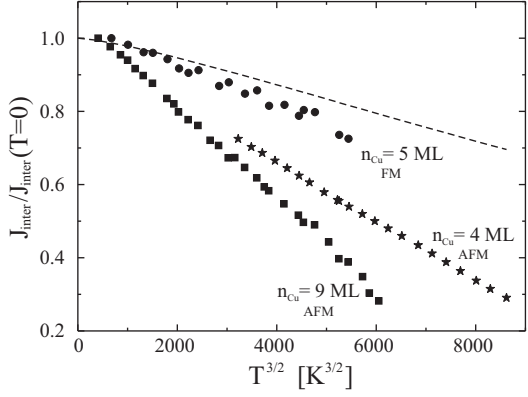}
\caption{Temperature dependence of $J_\text{inter}$ normalized to
$J_\text{inter}(T$=0) for the trilayer IEC system
Ni$_7$/Cu$_n$/Co$_2$/Cu(001) for three different spacer thicknesses
$n_\text{Cu}$ \cite{dr-lenz}. The data for 9 ML (full squares) are taken from 
Ref. \onlinecite{LM02}. The dashed line corresponds solely to a spacer effect
contribution for the $n_\text{Cu}=4~$ML sample.}
\label{fig1}
\end{figure}

Figure \ref{fig1} shows FMR experimental evidence (evaluated with LL) that
the magnetic layer effect gives indeed a significant contribution. 
According to the theoretical works by Bruno \cite{Bruno95} and Edwards
\emph{et al.}~\cite{Ed91} the $T$-dependence should be more pronounced for
a higher number of
spacer layers while it should be independent of the spacer thickness
within the spacer theory proposed in Ref.~\cite{Albu96}. The observed
stronger
decrease of the 4-ML-spacer (AFM coupled) system as compared to the
5-ML-spacer (FM coupled) system suggests that there are other than just
spacer mechanisms at work.
Moreover one sees a large difference between the measured
$J_\text{inter}$-curve of the 4-ML-spacer film and the theoretical curve
according to the spacer
theory suggested in Ref.~\cite{Bruno95}.

Ferromagnetic resonance is a powerful tool to investigate the IEC
experimentally \cite{LB03}. This technique probes the uniform spin wave mode
$\omega(\mathbf{q}=0)$ of a magnetic sample. An external field
$\mathbf{B}_{0}$ is tuned for a given probe frequency
$\nu_\text{hf}=\omega(\mathbf{q}=0)/2\pi$  until resonance
occurs at $B_\text{res}(T,\theta_{B_{0}})$, with $\theta_{B_{0}}$ being
the angle between the magnetic field and the normal to the film plane.
This resonance field
$B_\text{res}(T,\theta_{B_{0}})$ at which uniform
$(\mathbf{q}\!=\!0)$-spin wave modes with the energy $E^\text{SW}=h
\nu_\text{hf}$ of the probing microwaves are excited in the
magnetic system is the crucial quantity to connect experiment and theory.
With the method we propose it is possible to compute the spin wave energy 
$E^\text{SW}(\mathbf{q}=0,\mathbf{B}_{0},T)$ and read off the resonance
field given the (experimental) parameter $\nu_\text{hf}$, which was set to
$9~$GHz in our calculations \cite{SKN05b}.
For further details about the experimental procedure we refer to
Ref.~\cite{LB03} and give here a short outline of the experimental
procedure. After the preparation of
the Cu$_n$/Co$_2$/Cu(001) single film already capped with $n$ Cu spacer
layers, the angular dependence of the resonance field
$B^\text{Co}_\text{res}(T,\theta_{B_0})$ was determined at two
temperatures (55~K and 294~K). Then
the temperature dependence of the resonance field
$B^\text{Co}_\text{res}(T,\theta_{B_0})$ with $T \in
[55~\text{K},\approx\!300~\text{K}]$ at $\theta_{B_0}=90^{\circ}$ was
measured.
The same was carried out after the deposition of the top nickel layer for
the coupled system Ni$_7$/Cu$_9$/Co$_2$/Cu(001) and for comparison also
for the single film system Ni$_7$/Cu(001).

\begin{figure}[t]
\includegraphics[width=0.8\linewidth]{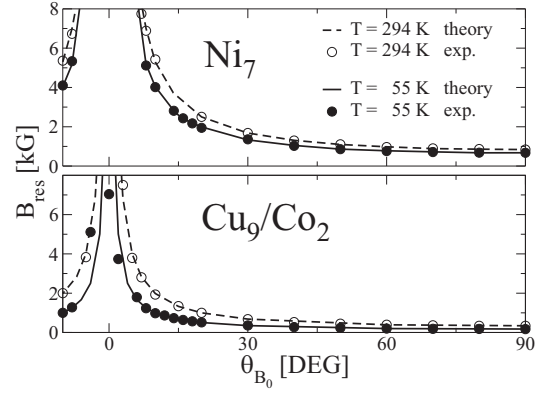}
\caption{Resonance field for Ni$_7$/Cu(001) and Cu$_9$/Co$_2$/Cu(001) as a
function of the orientation of the external magnetic field (circles).  The
angle $\theta_{B_{0}}$ is measured with respect to the film normal.  Lines
are theoretical fits, Ni: $S$=1, $K_2=3.0~\mu_\text{B}\text{kG}$,
$g_0=4.5~\mu_\text{B}\text{kG}$, $J=30$~meV, Co: $S$=2.5,
$K_2=-20.25~\mu_\text{B}\text{kG}$, $g_0=1.95~\mu_\text{B}\text{kG}$,
$J=4.1$~meV.}
\label{fig2}
\end{figure}
\begin{figure}[t]
\includegraphics[width=0.8\linewidth]{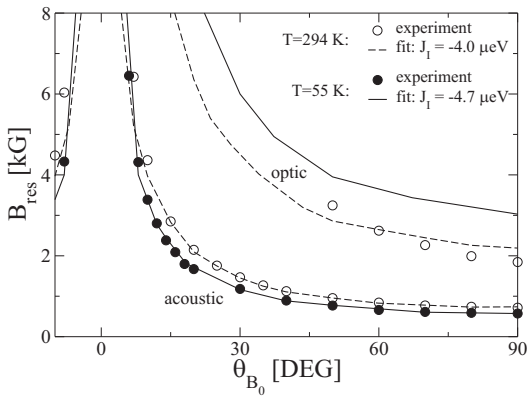}
\caption{Resonance field for the coupled system
Ni$_7$/Cu$_9$/Co$_2$/Cu(001). All parameters for the Ni and Co subsystems
are the same as in Fig.~\ref{fig2}.}
\label{fig3}
\end{figure}

\begin{figure}[t]
\includegraphics[width=0.8\linewidth]{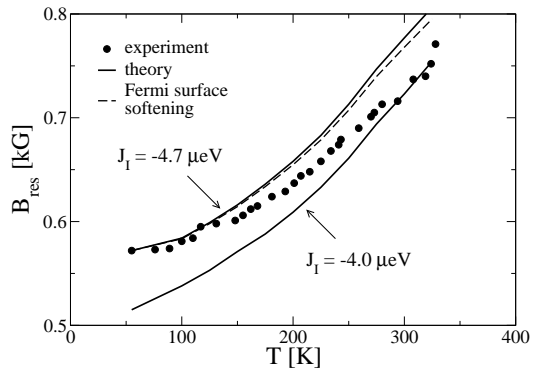}
\caption{Resonance field at $\theta_{B_0}=90^{\circ}$ as a function of
temperature for Ni$_7$/Cu$_9$/Co$_2$/Cu(001). Same parameters for Ni and
Co as in Figs.~\ref{fig2} and \ref{fig3}. Dashed line: the $T$-dependence 
due to the softening of the Cu Fermi surface \cite{Bruno95} was included.}
\label{fig4}
\end{figure}

\begin{figure}[t]
\includegraphics[width=0.8\linewidth]{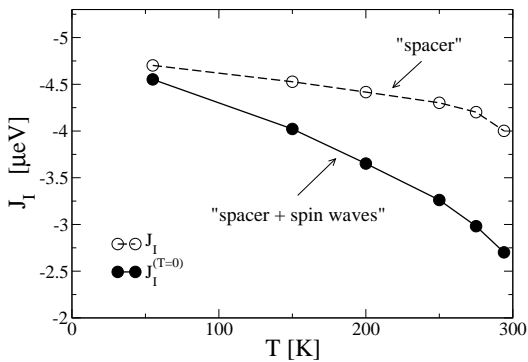}
\caption{Temperature dependence of interlayer exchange coupling in
Ni$_7$/Cu$_9$/Co$_2$/Cu(001). Symbols stem from fits of the experimental
resonance field at
$\theta=90^{\circ}$ at the given temperatures. Open circles: $T$-dependent
calculation. Filled circles: calculation at $T=0$. Lines are guides to the
eye.}
\label{fig5}
\end{figure}

The Hamiltonian of the effective two-layer system we use to describe IEC
reads
\begin{eqnarray}
H = &-&\sum_{<ij>\alpha}J_{\alpha}\mathbf{S}_{i\alpha}\mathbf{S}_{j\alpha}
\label{hamilt}
 -
\sum_{i\alpha}g_{\alpha}\mu_\text{B}\mathbf{B}_0\mathbf{S}_{i\alpha}-\\\nonumber
 &-&\sum^{\alpha \not=
\beta}_{<ij>\alpha\beta}J_{I}\mathbf{S}_{i\alpha}\mathbf{S}_{j\beta} -
\sum_{i\alpha}K_{2\alpha}S_{i\alpha z}^2 +\\
 &+&
\sum_{ij\alpha}g_{0\alpha}\left(\frac{1}{r_{ij}^3}\mathbf{S}_{i\alpha}\mathbf{S}_{j\alpha}
-
\frac{3}{r_{ij}^5}(\mathbf{S}_{i\alpha}\mathbf{r}_{ij})(\mathbf{S}_{j\alpha}\mathbf{r}_{ij})\right)\nonumber
\end{eqnarray}
The first term describes nearest-neighbor exchange coupling $J_{\alpha}$
between spin moments $\mathbf{S}_{i\alpha}$ and $\mathbf{S}_{j\alpha}$ at
sites $i$ and $j$
within the same layer $\alpha$ (=1,2).
The film thickness is implicitly included in the parameters $J_{\alpha}$
which are chosen such that the magnetic moments of the two monolayers described in (2)
equal that of the considered
multilayer films at room temperature, respectively ($T_\text{C}^{\text{Ni}_{7}}=410~$K,
 $T_\text{C}^{\text{Co}_{2}}= 400~$K).
The second term contains an external magnetic field
$\mathbf{B}_0$ in arbitrary direction with the spectroscopic splitting
factors $g_{\alpha}$ (taken as 2.2 for Ni and Co) and the Bohr magneton
$\mu_\text{B}$. The spacer
layer is not considered explicitly.
The IEC mediated by the spacer is described by the parameter $J_I$ (third
term). The fourth and fifth term represent second order lattice anisotropy
and dipolar
interaction, the latter leading to shape anisotropy. $K_{2\alpha}$ and
$g_{0\alpha}$ are microscopic anisotropy parameters, $S_{i\alpha z}$ is
the $z$-component of
$\mathbf{S}_{i\alpha}$ and is perpendicular to the film plane, $r_{ij}$ is
the distance between sites $i$ and $j$, and dipolar coupling across the
spacer
is neglected. The shape anisotropy favors in-plane orientation of the
magnetization, the lattice anisotropy in-plane ($K_{2\alpha}<0$) or
out-of-plane
($K_{2\alpha}>0$) orientation. Our Hamiltonian is similar to that used in
Ref.~\cite{Jens04} for the investigation of magnetic reorientation and
anisotropy.
We also point out that the local-moment aspect of inherently itinerant
magnets like Ni and Co has been demonstrated before: we mention here 
Heisenberg-like spin wave dispersions of an itinerant band model
\cite{BG01}, a high degree of the $d$-band moment localization ($>90\%$
for $0<T<T_\text{C}$) in an LDA + many-body approach to bulk Ni
\cite{Nol89}, and the successful employment of the Heisenberg model with
\emph{ab-initio} exchange parameters for the calculation of the Curie
temperature of Ni \cite{Bruno01}.

We summarize shortly how we solved (\ref{hamilt}) for the single-magnon
Green function (GF) $\langle \langle S_{i\alpha}^{+};S_{j\beta}^{-}\rangle
\rangle$
and refer to Refs.~\cite{SKN05a,SKN05b,PPS05} for details. In order to
treat the anisotropy terms properly we rotate the coordinate system
self-consistently
such that the new z-axis is parallel to the direction of the
magnetization. In the subsequent equation of motion approach it is
important to include all four
combinations of the GF $\langle \langle
S_{i\alpha}^{+,-};S_{j\beta}^{+,-}\rangle \rangle$ as was first suggested
in Ref.~\cite{PPS05}. This ensures correct
 softening properties of the uniform spin wave mode, namely its vanishing
 at the reorientation field for $K_{2}>0$ and up to a critical field strength for
 $K_{2}<0$, respectively.
Higher order GFs stemming from the Heisenberg exchange and from the
dipolar interaction are decoupled using the Tyablikov (RPA) method. In the
case of
the dipole-dipole interaction only the uniform (${\bf q}\rightarrow 0$)
contribution is considered as the nonuniform terms are negligible compared
to contributions
from the much larger Heisenberg exchange.
For the local $K_{2}$-terms an RPA decoupling fails. It was shown in
Ref.~\cite{SKN05a} that the Anderson-Callen decoupling yields quantitative
agreement
with QMC results for the field induced magnetic reorientation transition.
This is also the case for our improved theory which combines the proposal
of
Ref.~\cite{SKN05a} with the multiple GF approach by Pini \emph{et
al.}~\cite{PPS05}.
Different from the theoretical treatment based on the Landau-Lifshitz
equations, now with the quantum Heisenberg model a
self-consistent evaluation of the magnetization and the spin wave
excitation spectrum $E^\text{SW}({\bf q},\mathbf{B}_{0},T)$ is carried out.

In our theory the anisotropies $K_{2\alpha}$ and $g_{0\alpha}$ influence
the system solely via the {\em effective anisotropy field} given by the 
temperature-dependent term
\begin{eqnarray}
\label{Keff}
{\tilde K}_{2\alpha}(T)= \langle S_{z\alpha}\rangle(T)
\left(2K_{2\alpha}C_{\alpha}(T)-Dg_{0\alpha}\right)\;,\\
C_{\alpha}(T)=1-\frac{(S(S+1)-\langle S_{z\alpha}^2\rangle(T))}{2S2}\;.
\end{eqnarray}
Here the $T$-independent quantity $D$ is some number depending on the
lattice geometry. Note that our effective anisotropy field ${\tilde
K}_{2\alpha}$ corresponds
to the quantity $M_\text{eff}$ commonly used within a Landau-Lifshitz
description of FMR experiments \cite{LB03}. Furthermore we exploit 
$\langle S_{z\alpha}^2\rangle(T) = S(S+1)- \langle S_{z\alpha} \rangle(T)
[1+2\varphi_{\alpha}(T)]$ with the average magnon occupation number
$\varphi_{\alpha}$
\cite{SKN05b}.
The $T$-dependence of ${\tilde K}_{2\alpha}$ goes beyond a mere
proportionality to $\langle S_{z\alpha}\rangle(T)$ due to the expectation
value $\langle S_{z\alpha}^2\rangle(T)$.

Figure \ref{fig2} shows the comparison between the
$B_\text{res}(\theta_{B_{0}})$-curves from theory and experimental data
for a Cu$_9$/Co$_2$/Cu(001) and a Ni$_7$/Cu(001)
film system at two different temperatures. In both cases the effective
anisotropy favors the magnetization to lie within the film plane. There is
quite good agreement
at both temperatures over the whole range of angles $\theta_{B_{0}}$ for
both films.

It is important to note that for a fixed temperature the $B_\text{res}(\theta_{B_{0}})$ data alone do not allow one to determine the 
anisotropy parameters $K_{2\alpha}$ and $g_{0\alpha}$ independently of each other [see Eq.~(\ref{Keff})]. But taking into account both 
the $B_\text{res}(\theta_{B_{0}})$ data \emph{and} $B_\text{res}(T)$ at fixed $\theta_{B_0}$=90$^\circ$ the parameters can be separated 
because of the temperature dependent factor $C_{\alpha}(T)$ in Eq.~(\ref{Keff}). Indeed it is still possible to accurately fit the
experimental results with one set of ($T$-independent) parameters ($K_{2\alpha},~g_{0\alpha}$) for Ni and Co, respectively, over the whole
temperature range.
Furthermore in both cases the values of $g_{0\alpha}S$ lie slightly above
the result of an explicit evaluation of this quantity assuming point-like
dipoles on the
lattice sites for the given geometry \cite{LB03}
($g^{*}_{0}S=3.81\mu_\text{B}$kG). The conclusion we draw is that the
$T$-dependence of the effective magnetic
anisotropy is solely due to spin wave excitations which manifest
themselves in the $T$-dependence of (\ref{Keff}) rather than due to
$T$-dependent
$K_{2\alpha},~g_{0\alpha}$ which would effectively describe nonmagnonic
sources as thermal expansion or phononic interactions. This means that we
only have one fit
parameter, $J_I$, for the IEC system.

Figure \ref{fig3} shows the resonance fields for such a film system. {The
two branches at a given $T$ correspond to the acoustic (lower field branch) and
optical
(high field branch) excitation modes, respectively. The optical mode lying above the acoustic indicates 
antiferromagnetic coupling between Ni and Co.} The missing experimental
data for the optical
mode at $T=55~$K are due to a low intensity (oscillator strength) of the
resonance
signal \cite{LB03}. All parameters except for $J_I$ stem from the fits of
the single (uncoupled) Ni and Co films in Fig.~\ref{fig2}.

In addition we carried out fits at a fixed angle $\theta=90^{\circ}$ for a
variety of temperatures, the results of which are shown in
Fig.~\ref{fig4}. The variation of $J_{I}$ with $T$ represents the spacer contributions. 
From comparison with the results obtained by additionally taking Fermi surface softening 
into account we conclude that sources other than that dominate the spacer contribution, 
presumably the $T$-dependence of the reflection coefficients.  

In order to separate quantitatively the spacer and the magnetic
$T$-dependence of the IEC we fitted in a second step the same
$B_\text{res}$-data but setting the magnon occupation
number in our model artificially to zero, i.e., we ``switched off'' the
magnons.
The whole temperature dependence of the IEC \emph{including} magnon
excitations is then accounted for by the
fit parameter $J_{I}^{(0)}$. From the comparison between both resulting
curves of $J_{I}(T)$ and $J_{I}^{(0)}(T)$ in Fig.~\ref{fig5} one can read
off the weight of
each contribution to the $T$-dependence of the IEC. One concludes that the
magnonic part ($\approx 75\%$) exceeds the spacer part ($\approx
25\%$) significantly, however the latter is not negligible.

As for the variation of both contributions with the number of magnetic and
non-magnetic layers one can state the following general trend: the fewer
spacer layers and the thinner the magnetic part (i.e. the smaller
$T_{C}$), the more important spin waves become for IEC. Of course, for
bulk-like systems of Fe or Co with $T^{\rm bulk}_{C}>1000~$K the
magnon part will be (possibly much) less significant at room temperature or less. 
However, for $T_{C}$ below $\approx 700$ K, as for thin nano-film systems or 
Ni films or manganite multilayers \cite{Mn0105}, the spin wave contribution 
to the IEC temperature dependence becomes dominant.

In conclusion we proposed a novel approach that combines a quantum
theoretical treatment of the magnetic properties of multilayer film
systems with FMR experiments
on prototype Ni$_7$/Cu$_n$/Co$_2$/Cu(001) interlayer exchange coupled
films. Employing a microscopic theory based on an extended Heisenberg
Hamiltonian we studied the
temperature dependence of the IEC by fitting experimental data on the FMR
resonance field in two steps, first with $T$-dependent calculations and
then with $T=0$;
thus, ``switching off'' magnon excitations in the system. The result is
that magnon excitations are most important ($\approx 75\%$) for the
temperature dependence of
the interlayer exchange coupling. Band structure effects (e.g. smearing of
Fermi surface) are present but of minor importance. As a result we note
that a theoretical treatment of IEC at $T=0$ is of limited use when
interpreting the IEC of real multilayers at temperatures relevant for
technological applications.\\
Financial support by the DFG (Sfb 290) is gratefully acknowledged.

\end{document}